\documentclass[%
reprint,
superscriptaddress,
amsmath,amssymb,
aps,
]{revtex4-2}

\usepackage{subcaption}
\usepackage{siunitx}
\usepackage[utf8]{inputenc}
\usepackage[T1]{fontenc}
\usepackage{pgf}
\usepackage{comment}
\usepackage{graphicx}% Include figure files
\usepackage{dcolumn}% Align table columns on decimal point
\usepackage{bm}% bold math
\usepackage{amsfonts,amsmath,amsthm}
\usepackage{xcolor}
\usepackage{ulem}

\usepackage{xcolor}
\usepackage{soul}

\definecolor{myorange}{RGB}{255,220,120}
\sethlcolor{myorange}

\begin{document}
	
	\preprint{APS/123-QED}
	
	\title{Impact of light shift inhomogeneities on the contrast of light pulse atom interferometers}% Force line breaks with \\
	\author{Maxime Pesche}
    \affiliation{LTE, Observatoire de Paris, Université PSL, Sorbonne Université, Université de Lille, LNE, CNRS, 61 avenue de l’Observatoire, 75014 Paris, France}
		\author{Diego Alejandro Lancheros Naranjo}
    \affiliation{LTE, Observatoire de Paris, Université PSL, Sorbonne Université, Université de Lille, LNE, CNRS, 61 avenue de l’Observatoire, 75014 Paris, France}
    \author{Rayan Si-Ahmed} 
    \affiliation{LTE, Observatoire de Paris, Université PSL, Sorbonne Université, Université de Lille, LNE, CNRS, 61 avenue de l’Observatoire, 75014 Paris, France}
     \affiliation{Laboratoire Kastler Brossel, Sorbonne Université, CNRS, ENS-Université PSL, Collège de France, 75005 Paris, France}
    \author{Gabriel Ducasse}
	\affiliation{LTE, Observatoire de Paris, Université PSL, Sorbonne Université, Université de Lille, LNE, CNRS, 61 avenue de l’Observatoire, 75014 Paris, France}
	\author{Franck Pereira dos Santos}
	\affiliation{LTE, Observatoire de Paris, Université PSL, Sorbonne Université, Université de Lille, LNE, CNRS, 61 avenue de l’Observatoire, 75014 Paris, France}
	\author{Sébastien Merlet}
	\affiliation{LTE, Observatoire de Paris, Université PSL, Sorbonne Université, Université de Lille, LNE, CNRS, 61 avenue de l’Observatoire, 75014 Paris, France}%Lines break automatically or can be forced with \\

	\date{\today}% It is always \today, today,
	
	\begin{abstract}
    %Atomic interferometry is today one of most promising way of sensing inertial quantities with great accuracy and sensibility. A wave-packet can be projected into a superposition of two coherent state by the use of light pulses, different by their energy level as well as their momenta. With a well chosen sequence of pulses, one can generate an interferometer with a phase that can be sensible sensible to the acceleration of gravity, its gradient, the Earth rotation rate or the Casimir-Polder effect. The sensitivity of such a device is partially determined by the contrast in the atomic population, which value is affected by many effects already well established, such as the inefficiency of the light pulses or the spontaneous emission. This article focus on the effect on the contrast of an inhomogeneous light beam in intensity, that cause a distribution of phase printed among the whole atomic cloud. Measurements with the cold atoms absolute gravimeter (CAG) from the LTE manifest very well this effect, and its dependency with the duration of the pulse and the spatial separation of the wavepackets during the mirror pulse. A numerical simulation of the whole setup match those result better than one order of magnitude.
    
    We study the loss of the contrast in an atom interferometer when increasing the duration of Raman mirror pulses, and find the contrast decay rate to increase with the interferometer duration. We attribute this effect to the transverse spatial fluctuations of the intensity across the Raman beams, and to the dephasing induced by the associated light shifts inhomogeneities. Simulations based on the propagation of noisy synthetic Raman beams show that the contrast decay rate increases with the distance between the wavepackets at the mirror pulse, before reaching an asymptotic limit when intensity fluctuations between the two wavepackets become completely decorrelated. Finally, simulations based on the propagation of Raman beams having their measured intensity fluctuations predict contrast loss rates consistent with our measurements, confirming the detrimental role for the interferometer contrast played by intensity fluctuations across the interferometer laser beams.     
    
	\end{abstract}
	
	\maketitle
	
	%\tableofcontents

	\section{Introduction}
	
	In quantum sensors based on atom interferometry, the measurement stability is bound by a standard quantum limit set by the quantum projection noise \cite{Itano93,Santarelli99}. It is related to the projective measurement at the detection of the atomic populations in the output ports of the interferometer. For a given number of atoms $N$, this leads to a phase noise $\sigma_{\Phi} = \frac{1}{C\sqrt{N}}$, where $C$ is the contrast of the interferometer fringes, defined as its peak-to-peak amplitude. When working though with laser-cooled atomic samples at microkelvin temperatures, which contain of the order of $10^6-10^7$ detected atoms, the stability is in many cases actually not limited by detection noise, but rather by phase noise related to vibrations of the reference mirror used to retro-reflect the interferometer lasers \cite{Peters01,LeGouet08}. In this regime of large numbers of detected atoms, a limited contrast, provided it is not too low, is not of major concern. 
	
	However, using even colder atomic sources such as produced via evaporative cooling appears desirable to reduce systematics \cite{Karcher18,Zhang21}. But, this comes at the price of a drastic reduction in atom number, by about two orders of magnitude, which can make detection noise become a significant contribution to the instability of the measurement. Moreover, vibration noise can be efficiently suppressed in differential measurements \cite{Gustavson00,McGuirk02,Canuel06,Tackmann12}, as well as other common mode noise sources, or averaged thanks to joint measurements \cite{Savoie18}, allowing to reach the stability limit set by the quantum projection noise and the finite contrast \cite{Gauguet09,Sorrentino14,Janvier22}. The optimization of the stability when using ultracold atoms and/or differential configurations thus demands a good understanding of the impact of the different sources of contrast loss in an atom interferometer. 
    
This motivated numerous studies of sources of contrast loss in atom interferometers. A prime source for this degradation is the imperfect fidelity of the beamsplitters, limited on one hand by the velocity selectivity of the beamsplitters and the temperature of the atoms and on the other hand by inhomogeneities in the coupling due to the thermal expansion of the source across the laser beams \cite{Hu2017,Yang2023}. The use of ultracold atoms \cite{Zhang2021} and flat-top beams \cite{Mielec2018} allows mitigating the impact of these effects. In this paper, we report on a study performed on an atom gravimeter based on Raman beamsplitters of the impact of a different effect, also due to intensity inhomogeneities in the Raman beams, but leading to dephasing through inhomogeneous light shifts. Here, the decorrelation of the transverse intensity profiles along the propagation of the lasers plays a major role, so that, by contrast with previous works which consider these profiles to be identical at all pulses, one needs here to account for their variation along the direction of the laser field.
	
	The structure of the paper is as follows. We start by introducing our sensor, an absolute cold atom gravimeter (CAG) and by reviewing different sources of contrast loss commonly considered in such a setup. Having introduced the light shifts induced by the Raman lasers, we show how they impact, beyond the sole phase of the interferometer, its contrast. We then present the measurement protocol we have used to highlight their contribution to contrast loss in our interferometer. Numerical simulations based on the propagation of intensity fluctuations in synthetic laser beams are performed to quantify the effect for different intensity noise spectra. Finally, we compare our measurements with the results of a simulation performed for our actual Raman beam intensity profile.

	\section{Experimental setup}
	
	In our cold atom gravimeter, described in detail in \cite{Louchet11}, $^{87}$Rb atoms are trapped in a MOT before being cooled down to a temperature of $2~\mu$K in a far detuned molasses. They are then released in free fall for about $\SI{200}{\milli\second}$. We realize the selection of the atoms in the $m_F=0$ state with the combination of a microwave and pusher pulses. Using a single collimator placed on top of the chamber, we generate a gaussian beam with a waist of $\SI{11}{\milli\meter}$, containing two Raman laser fields with crossed circular polarizations $\sigma_-$ and $\sigma_+$. The beam is directed downward to the atoms and reflected on a system composed of a mirror and a quarter wave-plate, leaving the upward beam with reversed circular polarizations. This configuration allows generating conterpropagating stimulated Raman transitions, which will be used to realize the atom interferometer \cite{Kasevich91b}. Prior to the interferometer, a Raman pulse allows to perform a velocity selection \cite{Kasevich91a}, corresponding to an effective temperature of $0.7~\mu$K. The selected atoms are then interrogated by a sequence of three $\pi/2-\pi-\pi/2$ Raman pulses separated by an evolution time $T=\SI{80}{\milli\second}$, that respectively separate, redirect and recombine the matter waves, forming a Mach-Zehnder interferometer, such as displayed on Figure \ref{mach_zehnder}. The populations in the two interferometer outputs $N_1$ and $N_2$ are finally measured by a state selective fluorescence detection method at the bottom of the chamber \cite{Santarelli99}. The whole sequence lasts for $\SI{380}{\milli\second}$.

	%\begin{figure}[h]
	%	\centering
	%	\input{figure/frange_T.pgf}
	%	\caption{Probability transition of the atomic interferometer in function of the frequency ramp. Each color represent a different value of $T$ : \SI{65}{\milli\second} in red, \SI{70}{\milli\second} in blue, \SI{75}{\milli\second} in green and \SI{80}{\milli\second} in black} 
	%	\label{frange_T}
	%\end{figure}
	
	% cette figure n'est plus vraiment d'actualité
	
	The transition probability of the interferometer $P~=~N_1~/~(N_1+N_2)$ depends on the phase difference $\Delta\Phi$ accumulated along the two paths. Taking into account the free fall of the atoms and assuming ideal plane wave Raman beams, we can write $\Delta\Phi = \vec k_\text{eff} . \vec g T^2$, where $\vec k_\text{eff}$ is the effective wavevector of the Raman transitions and $\vec g$ is the gravity acceleration \cite{Borde89}. To compensate for the Doppler shift induced by the free fall of the atoms, we introduce a linear chirp $\alpha$ to the angular frequency difference between the Raman lasers in order to stay at resonance. This adds a contribution to the phase of the interferometer, which finally is written as $\Delta\Phi = (\vec k_\text{eff} . \vec g - \alpha) T^2$. Moreover, this provides a simple way to control the interferometer phase, allowing to record fringe patterns out of which the contrast can be extracted. We find in our nominal measurement conditions ($T=80$ ms) a contrast of 43\%. 

The main cause of this loss of contrast lies in the imperfect fidelity of the Raman pulses. In our interferometer, efficiencies of beamsplitters and mirror pulses are affected, for instance, by spontaneous emission, given that the detuning of our Raman lasers is \SI{-0.9}{\giga\hertz} with respect to the D2 line of $^{87}$Rb. An evaluation of the contrast loss due to this effect then yields a reduction of $3~\%$ for our standard three-pulse interferometer. A second source of inhomogeneity arises from the thermal expansion of the atomic cloud during free fall, which leads to a reduction in the Rabi coupling for the outermost atoms. A radial expansion on the order of $\SI{1}{\milli\meter}$ is expected, which corresponds to approximately one tenth of the chosen waist of the Raman beams. This results in a contrast loss of about $4\%$ due to thermal expansion. A  more stringent limitation to pulse efficiencies arises from the distribution of atom vertical velocities, which induces significant inhomogeneous Doppler detunings, and reduces the efficiency of the mirror pulse down to approximately $67~\%$. We calculate the resulting loss of contrast due to these imperfect resonant transitions, considering all three pulses, to be close to $39~\%$.
    
Dephasing is another source of contrast degradation, with the visibility of the fringes being washed out by the dispersion of the interferometer phase. Known causes for this type of contrast loss are fictitious forces due to rotations, such as the Coriolis, Euler and centrifugal forces \cite{Lan2012,Darmagnac2024,Marquet2026} or gravity gradients \cite{Roura2014,Roura2017}, especially significant for long interferometer durations and onboard applications \cite{Trimeche2019,Kaczmarczuk2025,Beaufils2023}. In our case, with a sensor fixed in a laboratory that rotates because of the Earth rotation, Coriolis acceleration is dominating. Atoms with an initial velocity $\vec{v}$ in the east-west direction acquire a Sagnac phase shift given by $\Phi_{C} = k_{\text{eff}} T^{2} (2 \vec{\Omega}_{E} \wedge \vec{v})$, where $\vec{\Omega}_{E}$ is the Earth rotation rate. The phase dispersion $\sigma_{\Phi_{C}}$ associated with the transverse velocity spread for atoms at a temperature of $2~\mu$K leads to a maximum contrast $e^{-\sigma_{\Phi_{C}}^2/2}$ of 0.98, corresponding to a contrast loss of $2~\%$.

By combining these three effects, we expect a contrast of the interferometer of $58~\%$, larger than the experimentally measured contrast of $43~\%$. This difference may come from an underestimation of coupling inhomogeneities, arising for instance to an imperfect centering of the atomic source in the beam, or a residual transverse velocity, or the presence of intensity fluctuations at shorter spatial scales than the beam waist.
    
In this work, we identify and characterize an other source of contrast loss, which arises from light-shift inhomogeneities during the mirror pulse, due to intensity fluctuations experienced differently by the two partial wave packets. Specifically, we show that the mean light-shift induces an additional phase in the interferometer, usually considered null, whose spatial inhomogeneity leads to dephasing and thus contrast loss. This non-trivial effect depends on the evolution of the transverse intensity profile of the Raman beams along their propagation and, consequently, on the spatial separation between the atomic wave packets.
    
	\section{Lightshift inhomogeneities at the mirror pulse}
	
	The Raman lasers, being out of resonance, induce a dynamic Stark energy shift on the levels they couple, here the two hyperfine ground states, denoted as $|f\rangle$ and $|e\rangle$ respectively, leading to parasitic contributions to the interferometer phase \cite{Weiss94}. In particular, a bias in the interferometer phase shows up, related to the differential light shift $\delta_\text{diff} = \delta_f - \delta_e$. This interferometer phase shift is written as:
	\begin{equation}
		\Delta \Phi_\text{diff} = \theta_1 - \theta_3
	\end{equation}
	where
	\begin{equation}
		\theta_i = -\arctan\!\left[\frac{\delta-\delta_{\mathrm{diff}}}{\Omega_R}\tan\!\left(\frac{\Omega_R \tau}{2}\right)\right],\quad {i= 1,3;}
	\end{equation}
	with $\Omega_R = \sqrt{ \Omega_\text{eff}^2 + (\delta-\delta_\text{diff})^2}$ the generalized Rabi frequency, and $\delta$ the Raman laser detuning. 

	Because of the ballistic expansion of the atomic cloud, the atoms explore regions of different intensities, and thus light shifts, over the duration of the interferometer. In particular, the radial positions of an atom at the first and third pulses in the Raman beam are different, so that $\theta_1$ and $\theta_3$ may be different. This effect can however be considerably mitigated by adjusting the power ratio of the two Raman lasers. Indeed, for a specific ratio of the intensities in the two Raman fields, of about 1.7 at our detuning, the differential light shift $\delta_\text{diff}$ is null, thus eliminating completely this effect at any position in the beam, provided that the two fields have exactly the same spatial mode. In addition, taking advantage that this phase shift does not depend on the direction $\vec k_\text{eff}$, while the gravity induced phase shift does, residual effects, related for instance to slow fluctuations of the intensity ratio between the fields, can in principle be rejected by averaging the gravity measurements obtained along the two directions of $\vec k_\text{eff}$ \cite{McGuirk02}.
	
	In addition, the interferometer phase may also be affected by the mean light shift $\delta_m = (\delta_f + \delta_e)/2$ \cite{Weiss94}. Since each partial wavepacket experiences a phase shift at each Raman pulse proportional to the light shift and the duration of the pulses, the resulting interferometer phase is written as:
	
	\begin{equation}
		\Delta \Phi_m = \sum_{i=1}^3 \left( \delta_{m,I}^i - \delta_{m,II}^i \right) \tau_i
	\end{equation}
	
	where $i$ is the pulse number, $\tau_i$ its duration and $\delta_{m,I}^i$ and $\delta_{m,II}^i$ are respectively the mean light shifts seen by the two wavepackets, on the arms $I$ and $II$ of the interferometer. One could argue that this contribution is null as the mean lightshift seen by each wavepacket is the same. If this holds for the first and the third Raman pulses, as they are at the same longitudinal position, for the second pulse, however, we must take into consideration the spatial separation between the wavepackets. This separation is linear with the free evolution time $T$ and reaches \SI{1}{\milli\meter} for our typical value of $T = \SI{80}{\milli\second}$. Since the two fields propagate, a fortiori with different wavevectors, intensities seen by each wavepacket can differ when their separation is large enough, resulting in an interferometer phase shift, linear with the duration of the second Raman pulse. This interferometer phase shift differs for each atom, since they all have a different trajectory into the Raman beam. We thus expect a distribution of interferometer phases over all of the atomic sample.
	
	In particular, if, despite being overlapped, the two Raman fields have different intensity fluctuations in the transverse profile of the laser beam, and if the dispersion of the resulting interferometer phase shifts is large enough, the contrast of the measurement could be reduced. As an example, for our typical second pulse duration of $32~\mu\text{s}$, a lightshift dispersion of \SI{1.5}{\kilo\hertz}, corresponding to one tenth of the Rabi coupling, causes a phase dispersion of \SI{0.3}{\radian} and a contrast loss of 4.5\%.

	\section{Experimental results}
	
	To highlight the effect of light shift inhomogeneities on the contrast, one shall observe the evolution of the contrast with the duration of the second Raman pulse, as the phase dispersion would increase linearly with this duration. A drawback of this method, though, lies in the presence of additional effects related to the Raman coupling, and separating the various effects from each other could be challenging. For instance, the change of pulse efficiency with its duration also affects the contrast. Despite these shortcomings, we will present at the end of this section such measurements, but we start by highlighting the impact of off resonant couplings using an alternative method.
	
	\subsection{Impact of a 4th pulse}
	
	We add a supplementary pulse in the standard 3 pulses interferometer sequence, between the first and the second ones. The resulting sequence is represented on the Figure \ref{mach_zehnder}, with standard pulses in red, and the additional pulse in blue. We detune the Raman lasers out of resonance during this additional pulse by offsetting the frequency of one of the two Raman lasers by \SI{2}{\mega\hertz}, preventing this pulse from coupling the two wavepackets, while leaving the light shift at play. We also present on the Figure \ref{mach_zehnder}, at the right, the measurements of the contrast as a function of the duration of the additional pulse $\tau_4$, that was varied from \SI{0} to $1000~\mu\text{s}$ %\SI{1000}{\micro\second}. 
    For these measurements, the evolution time $T$ is \SI{20}{\milli\second} and the non-resonant pulse is placed at a delay of $T_4 = \SI{10}{\milli\second}$ after the first pulse. The durations of the three pulses of the interferometer are $16-32-16~\mu\text{s}$ %$16-32-\SI{16}{\micro\second}$ 
    respectively and the intensity of the Raman beams is kept constant for the whole sequence. We observe a drastic contrast loss with $\tau_4$ that fits well with a Gaussian decay, displayed as a continuous red line, given by:

	\begin{figure}[h!]
        \centering
		\noindent\includegraphics[scale=0.7]{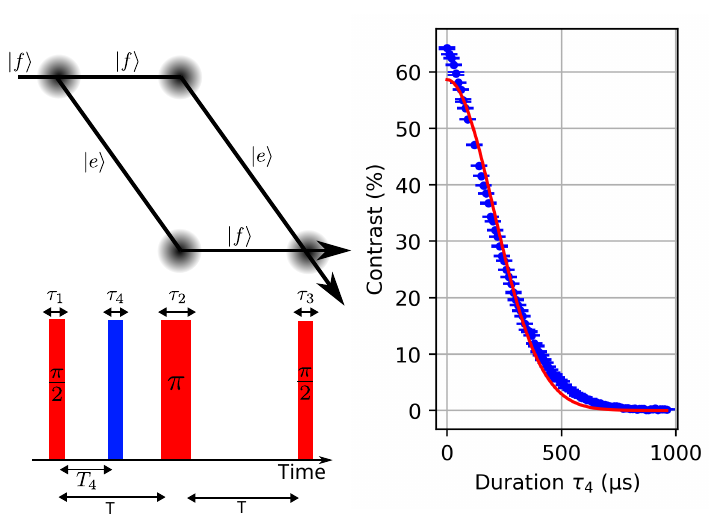}
		\caption{Left: modified interferometer sequence, with three resonant pulses in red, separated by the duration $T$, and an additional off-resonant pulse in blue, placed at a delay $T_4$ after the first pulse. Right: Measurement of the interferometer contrast versus the duration of the off-resonant pulse $\tau_4$, for $T=\SI{50}{\milli\second}$ and $T_4=\SI{40}{\milli\second}$ ms. The fit to the data with the equation \eqref{fit} is shown as a solid red line.} 
		\label{mach_zehnder}
	\end{figure}

	\begin{equation}
		C(\tau_4) = C_0 e^{- \frac {\tau_4^2}{\tau_c^2}}
		\label{fit}
	\end{equation}
	where $C_0$ is the maximum contrast and $\tau_c$ the characteristic decay time.
	
	%\begin{figure}[h]
	%	\centering
	%	\scalebox{0.6}{\input{figure/contraste_4th_desaccord.pgf}}
	%	\caption{Left: Contrast decay versus the duration of a non-resonant pulse placed at the delay of $T_4 = \SI{40}{\milli\second}$ after the first pulse, for different Raman detunings $\Delta$, ranging from \SI{0.4}{\giga\hertz} to \SI{1.2}{\giga\hertz}. Right: Corresponding characteristic decay times $\tau_C$ versus $\Delta$} 
	%	\label{contrast_detuning}
	%\end{figure}
	
	%\color{red}{la figure "contrast detuning" est à modifier pour la même avec le collimateur avec diaphragme, mais on a alors que deux valeurs de Delta. On pourrait la supprimer et juste dire dans le texte qu'on a vérifié que le temps caractéristique varie de X quand on varie le désaccord de X}
	%\color{black}
	
	To support the insight that the loss of contrast is related to dephasing by the light shifts, we measured the characteristic time $\tau_c$ as a function of the detuning $\Delta$ of the Raman lasers to the one-photon transition, and observed as expected a linear increase with $\Delta$. 
	
	%One can now study how the characteristic time $\tau_c$ depends on the experimental parameters, to confirm our insights. First, we measured the evolution of $\tau_C$ with respect to $\Delta$, the detuning of the Raman lasers to the one-photon transition. Since the Rabi coupling is inversely proportional to $\Delta$, we adapted for each detuning the duration of the 3 pulses of the interferometer to keep a $\pi/2 - \pi - \pi/2$ pulse sequence. For instance, at $\Delta = \SI{0.5}{\giga\hertz}$, twice smaller, the duration of the three pulses are doubled, leading to a sequence of $32-64-\SI{32}{\micro\second}$. We report in the left part of the figure \ref{contrast_detuning} the measurements of contrast decay versus the off-resonant pulse duration $\tau_4$ at different detunings, and in the right part, the corresponding characteristic decay times $\tau_C$ extracted from fits to the data. We observe that $\tau_C$ decreases linearly with decreasing detunings, in accordance with a lightshift effect, since the amplitude of the latter increases inversely proportional to $\Delta$.

	\begin{figure}[h]
		\centering
		\includegraphics{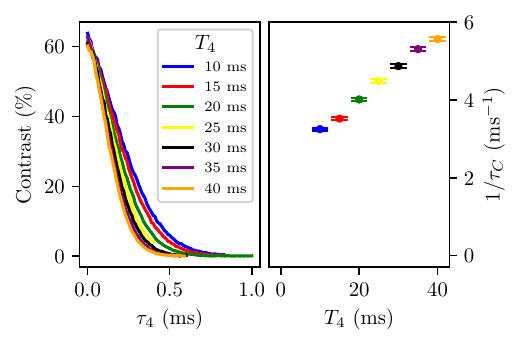}
		\caption{Left: Contrast measurement of a \SI{50}{\milli\second} interferometer with respect to the duration of a 4th pulse placed after a time $T_4$ after the first pulse. Right: inverse of the characteristic time with respect to the separation of the wave packet.} 
		\label{contrast_sep}
	\end{figure}

	%We then exploited the possibility to drive the off-resonant pulse at any position in the sequence, and for different free evolution times $T$. To highlight the importance of the wavepackets separation in the decay, we now use the protocol presented at the top of the figure \ref{contrast_sep}. Having fixed the evolution time $T$ at \SI{50}{\milli\second}, we now place the off-resonant pulse at variable delays $T_4$ after the first Raman pulse, ranging from \SI{1}{\milli\second} to \SI{40}{\milli\second}, which varies the separation $d$ between the two wavepackets. Note that we shift here the three resonant pulses of the interferometer in time, rather than the non-resonant pulse so that the position of the atoms in the Raman lasers at the off-resonant pulse is let unchanged.
	
	We then exploited the possibility to drive the off-resonant pulse at any position in the sequence, and for different free evolution times $T$. Having fixed the evolution time $T$ at \SI{50}{\milli\second}, we now place the first pulse at variable delays $T_4$ before the off-resonant Raman pulse, ranging from \SI{1}{\milli\second} to \SI{40}{\milli\second}, which varies the separation $d$ between the two wavepackets. Note that we shift here the three resonant pulses of the interferometer in time, rather than the non-resonant pulse so that the position of the atoms in the Raman lasers at the off-resonant pulse is left unchanged.
	
	Contrast decays versus $\tau_4$ and the corresponding $\tau_c$ versus $T_4$ are displayed on Figure \ref{contrast_sep}. We observe a linear variation of $1/\tau_c$ with $T_4$, consistent with a loss of contrast that depends on the spatial separation between the two arms of the interferometer. Note the non-zero (extrapolated) contrast loss at negligible separations, which cannot be explained solely by spontaneous emission, which induces for our parameters an exponential decay with a significantly lower rate of 0.6 ms$^{-1}$.
	
	\subsection{Impact of the second pulse}
	
\begin{figure}
	    \centering
	    \includegraphics[width=80mm]{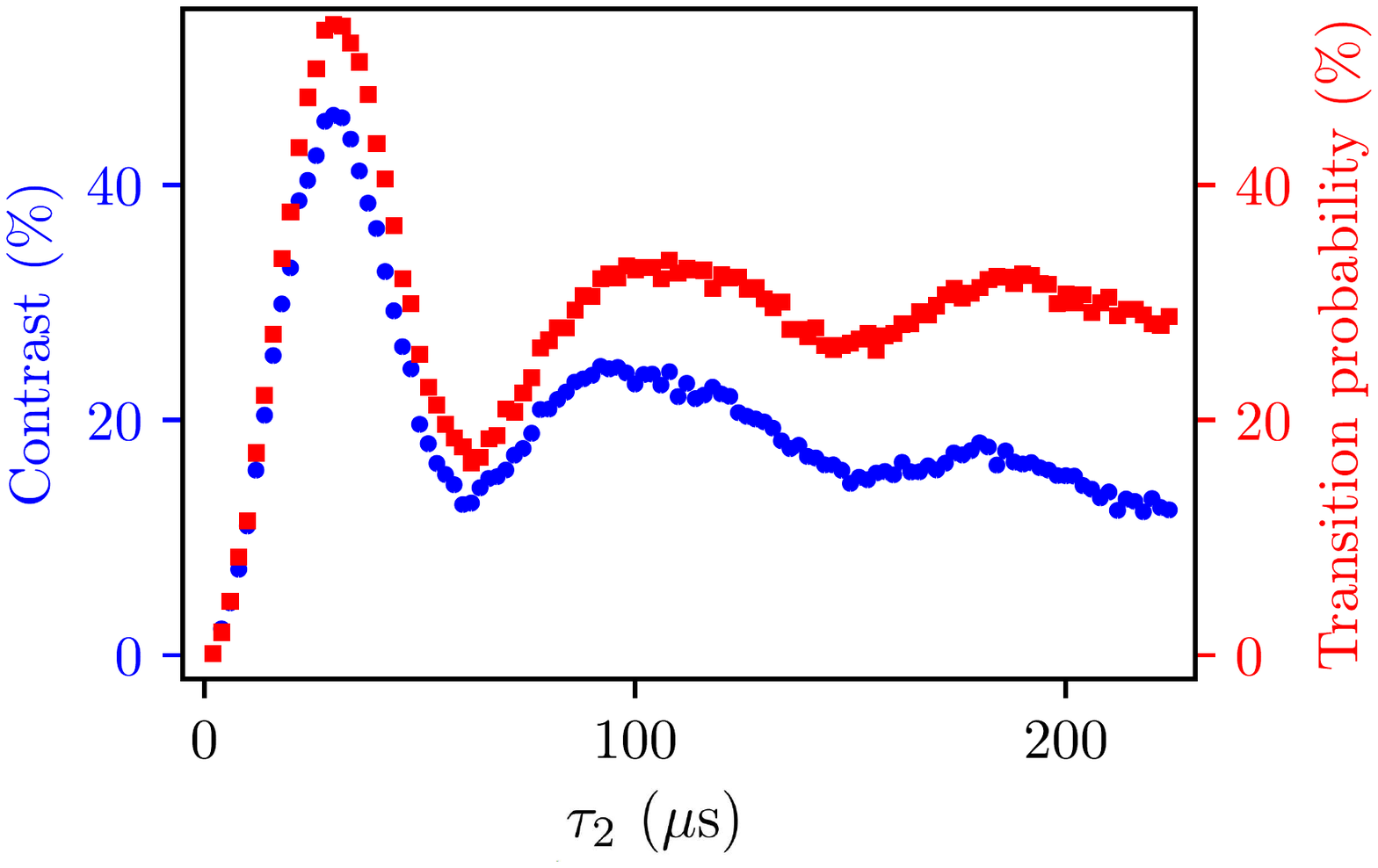}
	    \caption{Measurement of the contrast with respect to the duration of the second pulse (blue squares), as well as measured Rabi oscillations at the second pulse (red squares).}
	    \label{fig:2ndpulse1}
\end{figure}

\begin{figure}
	    \centering
	    \includegraphics{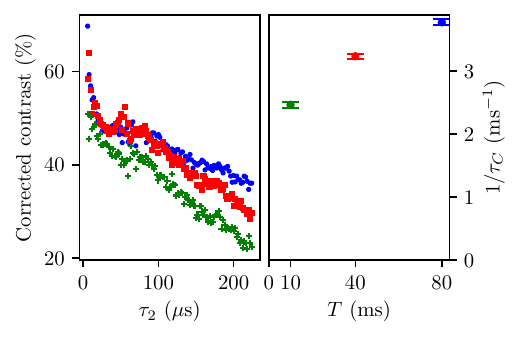}
	    \caption{Corrected contrast based on the 2nd pulse efficiency with respect to its duration for differents values of $T$. Right : Characteristic contrast decay times $\tau_c$ as a function of the evolution time $T$.}
	    \label{fig:2ndpulse2}
	\end{figure}

	Here, we perform measurements of the contrast as a function of the duration of the second pulse, keeping the duration of the first and third pulse unchanged and corresponding to $\pi/2$ pulses, for a Rabi frequency of 15~kHz, and for a free evolution time $T~=~80$~ms. Figure \ref{fig:2ndpulse1} displays as blue round symbols the measured contrast. It features oscillations due to the varying efficiency of the second pulse, which is displayed as red squares symbols by the Rabi oscillation measured at the second pulse. Dividing the measured contrast by the second pulse efficiency allows correcting for these oscillations, as displayed in Figure \ref{fig:2ndpulse2} at the left, where we recover an almost monotonous decay, which we attribute to light shift inhomogeneities. As above, we then extract the characteristic decay time $\tau_c$ by a Gaussian fit. We repeat this procedure for other values of $T$ (of 10~ms and 40~ms), and finally report on Figure \ref{fig:2ndpulse2} at the right all measured $\tau_c$ versus the corresponding separation distance between the two wavepackets at the second pulse. We find values of the order of the ones obtained above for the 4th pulse, but with a lower linear trend with the separation. This indicates that the observed decay at the second pulse does not originate from the resonant interaction between the atoms and the Raman beams, but rather from light shifts.
	
Finally, given the measured characteristic decay time for the nominal parameters of the interferometer $T=80$ ms, light shift inhomogeneities for the second pulse duration of $\tau_2=32\mu$s leads to a non negligible contrast loss about $2\%$. This is comparable to the contribution of spontaneous emission or dephasing due to Coriolis accelerations, even if it cannot not explain by itself the difference between the measured and expected contrasts.
    
	\section{Numerical simulation}
	
We now perform numerical simulations to study the impact on the contrast of transverse intensity variations in the laser beams,
    characterized by fluctuations at different spatial scales. We show that the propagation of the fields leads to increasing decorrelation of their intensity distributions, leading to light shift inhomogeneities, whether differential or mean, even at the ideal ratio of intensities in the two fields.
	
	We start by defining the Raman laser beam intensity and phase profile over a $1000~\times~1000$ grid with pixel size of \SI{7.6}{\micro \meter} in an initial plane. We choose uniform phase and intensity distributions, to which we add random intensity fluctuations of variable spatial frequency content. For that, we start by drawing random values in a normal distribution at each pixel of the grid. We then transform to Fourier space, where an exponential low pass filter is applied with an angular spatial frequency cutoff $k_c~=~1/\sigma$. Going back to real space, the laser beam exhibits fluctuations with a correlation length $\lambda_c=2\pi\sigma$. Last, we normalize these fluctuations to a given RMS value $\epsilon$. 
	
	Before addressing the effect of propagation on the differential intensity between the two Raman beams, we first characterize how a single beam with controlled transverse fluctuations evolves under free space propagation. Starting from an initial  flat intensity and phase profile, random transverse amplitude fluctuations with a prescribed spatial frequency content are applied and the resulting field is propagated over a distance $z$. 
	
\begin{figure}[h]
		\centering
		\includegraphics[width=1\linewidth]{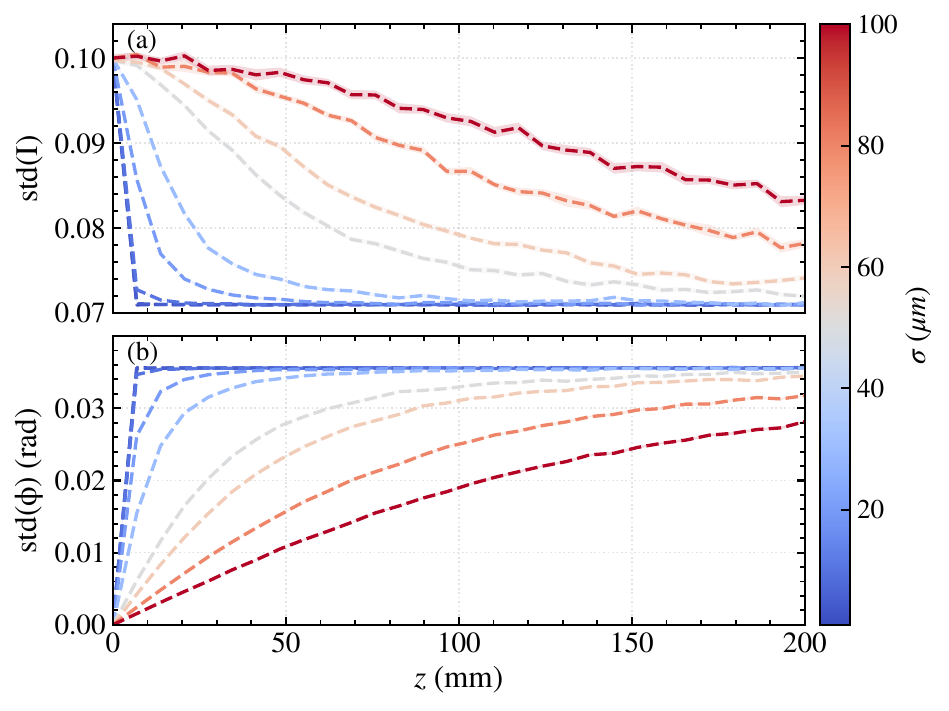}
		\caption{Standard deviation of the normalized intensity (a) and  standard deviation of the phase (b) as a function of the propagation distance, for $\epsilon=10\%$ and for different values of the cutoff parameter $\sigma$. Each point is averaged over several independent realizations of the initial random intensity pattern.} 
		\label{fig:std_I_and_phi_one_beam}
	\end{figure}

	The statistical properties of the propagated field are evaluated within a centered transverse window and characterized by two observables: the standard deviation of the normalized intensity and the circular standard deviation of the optical phase. Figure \ref{fig:std_I_and_phi_one_beam} shows the evolution of these quantities as a function of the propagation distance for $\epsilon=10\%$ and for different values of the cutoff parameter $\sigma$, each point being averaged over several independent realizations of the initial random pattern.
	
	For all values of the cutoff $\sigma$, the propagation leads to a rapid evolution of phase and intensity standard deviations, followed by a saturation regime. The initial transverse amplitude fluctuations are progressively converted into phase fluctuations through diffraction, while small-scale intensity modulations are redistributed over the transverse profile of the beam. As a result, the local intensity pattern loses memory of its initial configuration as the propagation distance increases.
	
	The characteristic propagation length over which this redistribution occurs strongly depends on the transverse correlation length of the fluctuations. Smaller values of $\sigma$, corresponding to shorter transverse correlation lengths and larger transverse spatial frequencies, lead to faster saturation. This behaviour reflects the quadratic dispersion relation of free space propagation, for which the accumulated phase scales as $k_\perp^2 z$, leading to a characteristic decorrelation length scaling as $\sigma^2$.
	
\begin{figure}[h]
		\centering
		\includegraphics[width=1\linewidth]{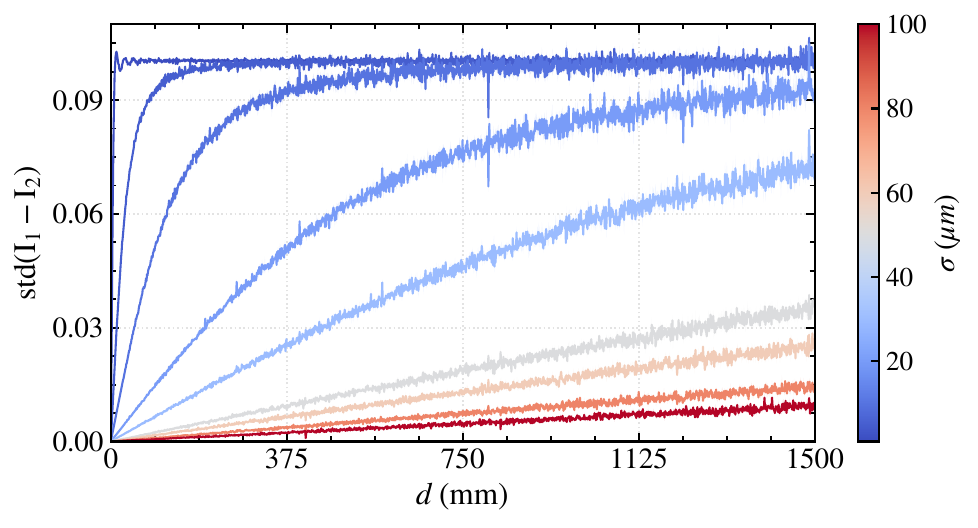}
		\caption{Standard deviation of the intensity difference between the two Raman lasers as a function of their common propagation distance} 
		\label{fig:xx}
	\end{figure}

	In a second simulation, we propagate the two Raman beams, assuming they have the same profile in the initial plane, but taking into account the difference in their frequencies of 6.8 GHz. We expect that the propagation leads to an increasing decorrelation between the intensity spatial fluctuations of the two beams. We quantify this effect by calculating the standard deviation of the difference of the normalized intensities in the same plane. Figure \ref{fig:xx} displays this standard deviation as a function of the propagating distance $z$, for different cutoffs $\sigma$ and $\epsilon=10\%$. As expected, standard deviations increase with distance, before reaching a plateau after propagation over distances of typically a few tens of meters. The plateau corresponds to a complete decorrelation between local intensity fluctuations in the two beams. Note that at large propagation distances, where the two Raman intensity patterns are fully decorrelated, residual fluctuations in the estimation of the standard deviations originate from the finite number  of transverse modes within the observation window, rather than from insufficient Monte Carlo averaging. The standard deviations actually scales as $z/\sigma^2$, so that smaller cuttoffs lead to faster saturation with the distance.  
	
	In a third simulation, we propagate one of the two beams and calculate the difference between the intensities in two distinct forward planes as a function of their separation.
	
	\begin{figure}[h]
		\centering
		\includegraphics[width=1\linewidth]{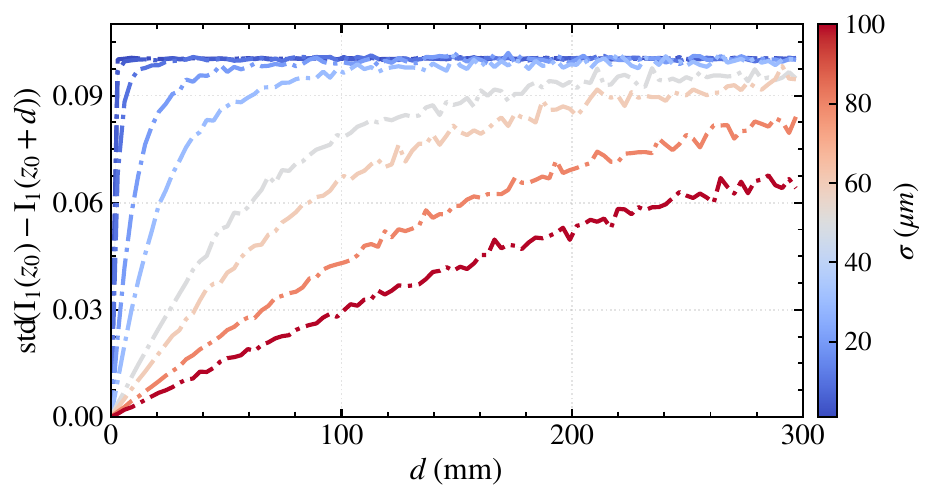}
		\caption{Standard deviation of the intensity difference between two different planes for one Raman beam} 
		\label{fig:xy}
	\end{figure}

Figure \ref{fig:xy} displays the standard deviation of this difference as a function of this separation distance $z$, for different cutoffs $\sigma$. We observe a similar behaviour as for the intensity difference between the two beams in the same plane, but with saturations reached after much shorter distances, of order of centimeters. 

These simulations show that the decorrelation is due mainly to the difference in propagation distance and to a smaller extent to the difference between the frequencies of the laser. It will induce fluctuations of the light shift across the Raman beams, and eventually dephasing. 

To evaluate these light shift fluctuations, we now calculate the intensity distributions of the Raman beams in two different planes separated by a distance $d$. This distance will correspond to the centers of the two wavepackets at the mirror pulse of the interferometer or at the non-resonant additional pulse we can add to the standard three pulse sequence. The two Raman beams exit from the same collimator, whose output plane is taken here as our initial plane, and are then reflected by the reference mirror. We thus have to calculate in each of the two planes, placed at distances $z$ and $z+d$ from the initial plane to the atoms, the intensities of four beams: two propagating over the distances $z$ and $z+d$ directly, and two propagating over the distance $2L-z$ and $2L-z-d$ since reflecting on the mirror, $L$ being the distance between the initial plane and the mirror. The corresponding eight normalized intensities are denoted by $I_i(u)$, where $i\in \{1,2\}$ stands for the two Raman frequencies, and $u\in \{z,2L-z,z+d,2L-z-d\}$ for the four possible propagation distances.
	
	\subsection{Effect of a non-resonant pulse}
	Using these intensities, we first calculate the light shifts experienced by the two internal states in the different interferometer arms, of relevance for the experiments with the additional non-resonant pulse. At our detuning of 900~MHz with respect to the $F'=1$ state, for a Rabi frequency of 15.4 kHz and a ratio between the intensities of the Raman lasers of $r~=~1.73$, we calculate light shifts (in units of kHz) of 
	
	\footnotesize
	\begin{align}
		\delta_f(z)/2\pi&=-12.4(I_1(z)+I_1(2L-z))-2.6(I_2(z)+I_2(2L-z))\\
		\delta_e(z)/2\pi&=2.1(I_1(z)+I_1(2L-z))-17.1(I_2(z)+I_2(2L-z))
	\end{align}
	\normalsize
	
	Note that the contributions of the individual beams are summed, neglecting any interference effects, which for our polarization configuration are in principle not present. Note as well that in the absence of intensity fluctuations ($I_i(u)=1$), the differential light shift is null. 
	
	We then compute the transverse standard deviation of the light-shifts difference experienced by the two separated wavepackets across the transverse profile of the beam $\Sigma~=~\sigma(\delta_f(z)~-~\delta_e(z+d))$, the corresponding phase dispersion $\sigma_\Phi~=~\Sigma~\tau_4$ induced by an additional non-resonant pulse of duration $\tau_4$, and characteristic decay time $\tau_c~=~\sqrt{2}/\Sigma$.     
	
\begin{figure}[h]
	\centering
	    %\begin{subfigure}{.25\textwidth}
	    %    \centering
	    %    \includegraphics[width=1.0\linewidth]{figure/SigmaVsd.png}
		%    %\caption{Relative intensity fluctuations}
		%    \label{fig:sub1}
	    %\end{subfigure}%
		%\begin{subfigure}{.25\textwidth}
		%	\centering
		%	\includegraphics[width=1.0\linewidth]{figure/SigmaVsparam.png}
		%	%\caption{Radial PSD of relative intensity fluctuations}
		%	\label{fig:sub2}
		%\end{subfigure}
		\includegraphics[width=1\linewidth]{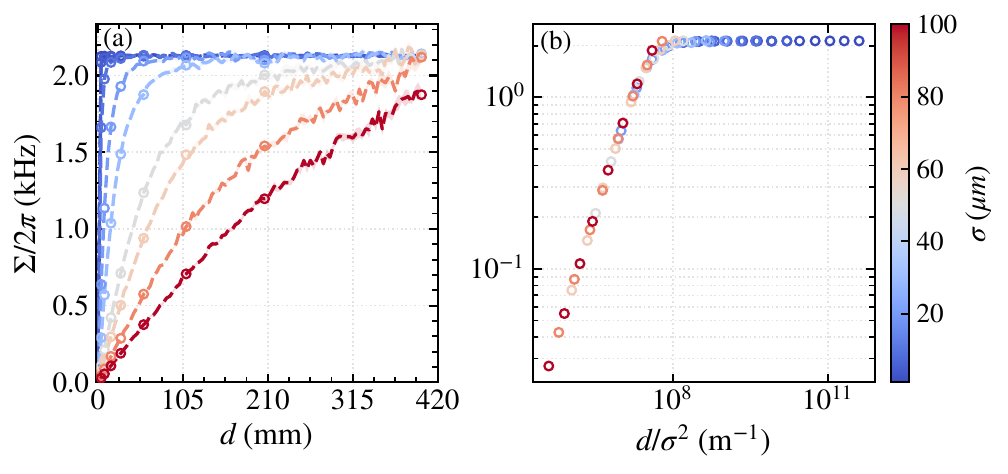}
		\caption{Standard deviation of the light shift difference as a function of the separation $d$  (left) and as a function of the normalized parameter $d/\sigma^2$ (right).}
		\label{fig:stdvsdsimu}
	\end{figure}

	Figure \ref{fig:stdvsdsimu} displays at the left the calculated values of $\Sigma/2\pi$, for different cutoffs $\sigma$ and for an amplitude $\epsilon~=~10\%$, as a function of the separation $d$ (left) and of the scaling parameter $d/\sigma^2$ (right).

$\Sigma$ is found to increase with the distance $d$ before saturating to an asymptotic value that corresponds to completely uncorrelated intensity fluctuations in the different laser beams. This value can be derived as following. Neglecting the difference of wavelengths between the two lasers, the light shift difference is written as
\begin{align*}
(\delta_f(z)-\delta_e(z+d))/2\pi&=15(I_1(z)+I_1(2L-z)\\
&-I_1(z+d)-I_1(2L-z-d))
\end{align*}
which for uncorrelated intensity fluctuations, leads to  \begin{equation}
	\Sigma/2\pi =15\times \sqrt{4} \epsilon/\sqrt{2} =2.1~\textrm{kHz}    
	\end{equation}  
The corresponding asymptotic value of the characteristic decay time amounts to $\tau_c=106~\mu$s, of the order of the measured values.

Finally, we find that $\Sigma$ follows the following scaling law
	\begin{equation}
	\frac{\epsilon^2}{\Sigma^2} = A + B \times \left(\frac{\sigma^2}{ d}\right)^2  
	\end{equation}
%    with $A = \SI{5.6e-9}{} \pm \SI{4e-13}{\per \hertz \squared}$ and $B = \SI{5.6e-4}{} \pm \SI{70}{\per\hertz\squared \per\meter\squared}$
    with $A~=~5.6\times10^{-4}$~(krad/s)$^{-2}$ and $B~=~5.6\times10^{11}$ (krad/s~m)$^{-2}$.

	\subsection{Effect at the mirror pulse}
	
	We now calculate the mean light shifts experienced by the two different interferometer arms, of relevance for the experiments where the duration of the second pulse is increased, for the same Rabi coupling of 15.4~kHz and ratio between the intensities of the Raman lasers $r~=~1.73$: 
	
	\begin{align}
		\delta_I&=(\delta_f(z)+\delta_e(z))/2\\
		\delta_{II}=&(\delta_f(z+d)+\delta_e(z+d))/2
	\end{align}
	
	As in the previous subsection, we then calculate the standard deviation of the difference of mean light shifts experienced by the two arms across the transverse profile of the beam $\Sigma_2=\sigma(\delta_I-\delta_{II})$, the corresponding phase dispersion $\sigma_\Phi=\Sigma_2 \tau_2$ induced by the mirror pulse of duration $\tau_2$, and characteristic decay time $\tau_{c,2}=\sqrt{2}/\Sigma_2$.     
	
The results we obtain for the second pulse are actually very similar to those of the fourth pulse, which stems from the small difference in the wavelength difference between the two Raman lasers. Indeed, one finds the same light shift differences in the two cases, if neglecting their wavelength difference. Therefore, for large separations $d$ and for rms intensity fluctuations at the atoms of $\epsilon/\sqrt{2}=7\%$, the phase dispersion due to the light shift at the second pulse reaches up to $\sigma_\Phi=0.42$ rad, which degrades the contrast by $9\%$.
    %if we neglect the difference between the wavelengths of the two Raman lasers, one finds the same effects for the 2nd pulse and for a 4th pulse with the same spatial separation. 

\subsection{Summary}
The simulations show that the inhomogeneity in the difference between the light shifts, either at the second or the forth pulse, and thus the dephasing they induce, increases with the separation between the wavepackets, up to a saturated value where intensity fluctuations in the different laser beams are completely decorrelated. Since the model spectra we have considered here do not necessarily correspond to a real laser beam, a quantitative comparison between the values we calculated in this section and the measured values presented above is not straightforward. For that comparison,  we will account in the next section for the spectrum of the very beam used for the measurements.

	\section{Laser beam profile}
	
\begin{figure}[h]
		\centering
		\includegraphics[scale=0.4]{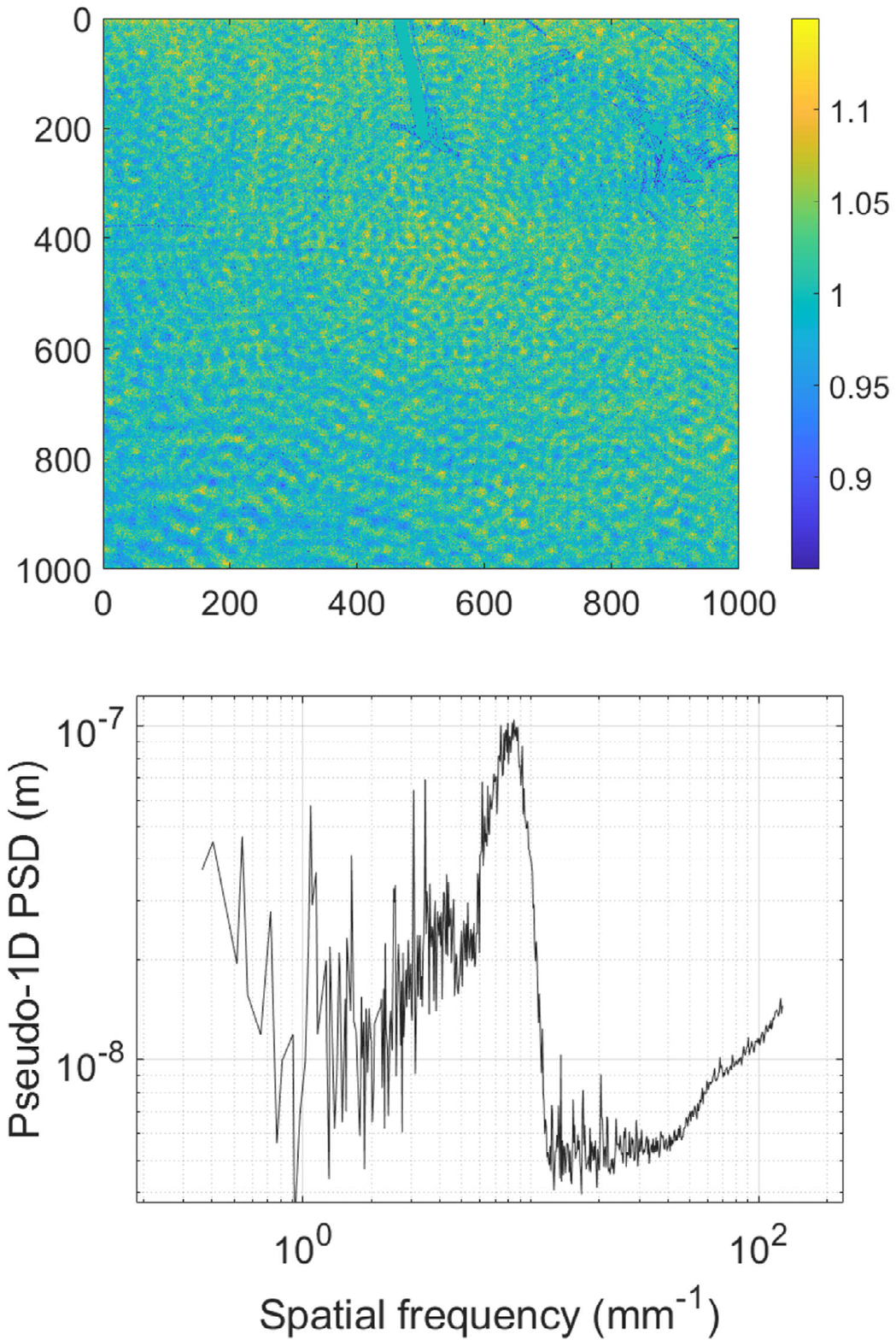}
		\caption{Relative intensity at the center of the Raman lasers and the corresponding pseudo-1D PSD.}
		\label{fig:test}
	\end{figure}

	Finally, we removed the Raman collimator outside the magnetic shields in which it is installed, and we measured the Raman laser intensity profile at its output with a CCD camera ($1088~\times~ 2048$~pixels of 5.5~$\mu$m size length). Since the size of the sensor is smaller than the clear aperture of the beam, we measure here the intensity at the center of the beam, but over an area that remains larger than the cloud size, even for the longest free fall times. We then select the central portion of the picture ($1000~\times~1000$~pixels) and normalize the intensity profile, which is displayed on Figure \ref{fig:test} at the top. 
	We also provide in Figure \ref{fig:test} at the bottom the corresponding pseudo-1D Power Spectral Density (PSD) of relative intensity fluctuations, which features a smoother and more realistic decrease with spatial frequency when compared to synthetic profiles with exponential filtering. 

	Similar to what was done with synthetic profiles, we then calculated out of its propagation the expected impact on the contrast.

\begin{figure}[h]
		\centering
		\includegraphics[width=1\linewidth]{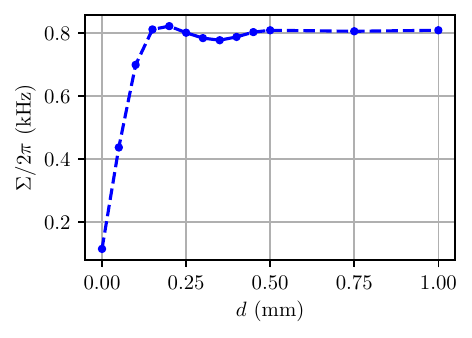}
		\caption{Standard deviation of the light shift difference as a function of the separation} 
		\label{fig:stdls}
	\end{figure}

	Figure \ref{fig:stdls} displays the corresponding standard deviations of $\Sigma/2\pi$, for separations between the wavepackets ranging from 0 to 1 mm. We observe a saturation of the effect, due to the rapid decorrelation induced by fluctuations of large spatial frequencies. Note that since the position of the atomic cloud with respect to the subpicture is not precisely known, we performed the calculation over the full size of the subpicture, rather than over the distribution of the atom cloud. We checked though that the dispersion calculated over the atomic cloud does not vary significantly with its position in the subpicture and remains close to the result obtained over the full surface of it.

	\begin{figure}[h]
		\centering
		\includegraphics[width=1\linewidth]{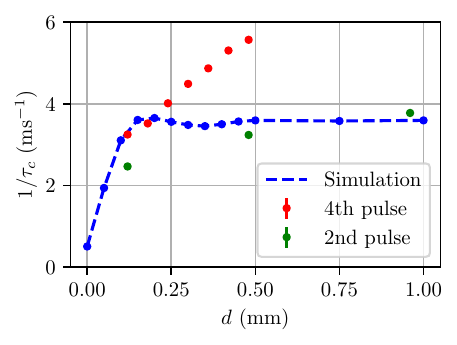}
		\caption{Characteristic contrast decay times as a function of the separation} 
		\label{fig:invtauc}
	\end{figure}

Finally, we compare in Figure \ref{fig:invtauc} the inverse of the calculated characteristic times with the same measured either with a 4th or directly at the 2nd pulse. We find a certain agreement between measurements and calculations, since they all show an increase with the wavepacket separation, and values of the same order of magnitude. Though the measurements do not clearly show a saturation, they do not either extrapolate to zero for null separation, which is consistent with the non-linear behaviour observed in the calculation. Differences between the calculated and measured values could be due to the impact of the many optical elements in the path of the Raman beam (one mirror outside the chamber, one input viewport and the retroreflecting optics inside the chamber, which consist in the combination of a quarter-wave plate and a mirror). In particular, dust on the optics outside the vacuum chamber could lead to significant degradation of the intensity profile.

	\section{Conclusion}
    
We highlight and quantify the impact of light shift inhomogeneities in the Raman beams on the contrast of an atom interferometer in a gravimeter configuration. We show their contribution to contrast loss can be comparable to other effects, such as the dephasing of Coriolis accelerations and the coupling inhomgeneities. In particular, we obtain evidence for the dependence of the effect on the separation of the wave-packets. Calculation of the dephasing induced by light shift inhomogeneities based on simulations of the propagation of the Raman beams show the role played by the decorrelation induced by the propagation of the Raman beams. 

In the case of the Raman beams used in the interferometer, we obtain a characteristic correlation length on the order of $100~\mu\text{m}$, sufficient to reach saturation in the characteristic contrast decay time. Improving the quality of the optical system, with larger correlation lengths and lower intensity noise at high spatial frequencies, would mitigate the impact of light shift inhomogeneities. Another strategy to reduce the impact of the light shifts would be to add blue detuned additional laser lines, so as to cancel the light shifts and thus their inhomogeneity \cite{Kovachy2015}.

Our study is another illustration of the detrimental impact of laser intensity fluctuations in atom interferometers based on light beamsplitters, also responsible for the modification of the photon momentum \cite{Bade2018,Gaudout2025} and the presence of dipole forces \cite{Kovachy2015}. Our quantitative analysis allows to set clear constraints on these intensity fluctuations, in addition to the phase wavefront aberrations that are more commonly taken into account. As a perspective, since phase and intensity are linked to the complex amplitude of the electric field, a more comprehensive treatment of the impact of the laser beams, where both phase and intensity fluctuations and their propagation are taken into account, will be necessary to properly provide specifications on the optical quality of the lasers of the interferometer. 
Finally, if we find the impact of light shift inhomogeneities to be relatively modest for our nominal 3-pulse interferometers, they could impose more severe limits when implementing multipulse sequences \cite{Mcguirk2000,Leveque2009,Butts2013,Dubetsky2023,Chen2024} or optimal control methods \cite{Saywell2020}, which demand to increase significantly the Rabi area of the mirror pulse.

\section{Acknowlegments}
	
The authors acknowledge financial support from the Agence Nationale de la Recherche through the TONICS project "ANR-21-CE47-0017", from
the Qu-Test project, which has received funding from
the European Union’s Horizon Europe research and
innovation program under agreement No 101113901 and from a government grant managed by the Agence Nationale de la Recherche under the Plan France 2030 through the QAFCA project “ANR-22-PETQ-0005”. This work has also been supported by Région Ile-de-France in the framework of DIM SIRTEQ through the project "TWAIN". M.P. acknowledges the support by the Labex First-TF, managed by the Agence Nationale de la Recherche under the reference ANR-10-LABX-48-01, through the project "GAUFR". The authors thank Pierre Cladé and Saïda Guellati-Khelifa for useful discussions and for sharing with us their numerical simulation tools.

\bibliography{ArticleContrastDecay}% Produces the bibliography via BibTeX.
		
	\end{document}